\documentclass[12pt]{article}
\usepackage{amssymb}
\usepackage{epsfig}
\usepackage[english]{babel}
\usepackage{float}
\usepackage{slashed}
\newcommand{\be}{\begin{equation}}
\newcommand{\ee}{\end{equation}}
\newcommand{\bea}{\begin{eqnarray}}
\newcommand{\eea}{\end{eqnarray}}

\def\p{\partial}
\def\pslash{\p\raise.3ex \hbox{\kern-.5em /}}
\def\delslash{\nabla\raise.3ex \hbox{\kern-.7em /}}

\begin{document}

\vskip 5cm

\begin{center}
\Large{ \textbf{Non-Hermitian ${\cal PT}$-symmetric relativistic
quantum theory in an intensive magnetic field }}
\end{center}
\vskip 0.5cm\begin{center} \Large{V.N.Rodionov}
\end{center}
\vskip 0.5cm
\begin{center}
{Plekhanov Russian University of Economics, Moscow, Russia,  \em
E-mail rodyvn@mail.ru}
\end{center}

\begin{center}

\abstract{We develop relativistic non-Hermitian quantum theory and
its application to neutrino physics in a strong magnetic field. It
is well known, that one of the fundamental postulates of quantum
theory is the requirement of Hermiticity of physical parameters.
This condition not only guarantees the reality of the eigenvalues
of Hamiltonian operators, but also implies the preservation of the
probabilities of the considered quantum processes. However as it
was shown relatively recently (Bender, Boettcher 1998),
Hermiticity is a sufficient but it is not a necessary condition.
It turned out that among non-Hermitian Hamiltonians it is possible
to allocate a number of such which have real energy spectra and
can ensure the development of systems over time with preserving
unitarity. This type of Hamiltonians includes  so-called
parity-time (${\cal PT}$) symmetric models which is already used
in various fields of modern physics. The most developed in this
respect are models, which used in the field of ${\cal
PT}$-symmetric optics, where for several years produced not only
theoretical but experimental studies.  }

\end{center}


\section{Introduction}

It is well known the Nobel Prize in Physics was awarded 2015
jointly to Takaaki Kajita and Arthur B. McDonald "for the
discovery of neutrino oscillations, which shows that neutrinos
have mass". This discovery has completely changed our
understanding of the innermost properties of matter and showed
that Standard Model (SM) cannot be the comprehensive theory of the
fundamental constituents of the Universe. Obviously, that of past
successes of the SM is so high that new models which is designed
for modifying SM, should contain practically all basic principles
lying in the basis of already existing theory. It is important to
note that the generalization of the notion of Hermiticity has a
strict quantum mechanical justification. Indeed it is known one of
the fundamental postulates of quantum theory is the requirement of
Hermiticity of physical parameters. This condition not only
guarantees the reality of the eigenvalues of Hamilton operators,
but also implies the preservation in time of the probabilities of
the considered quantum processes. However, as was shown relatively
recently \cite{ben}, Hermiticity is a sufficient but it is not a
necessary condition. It turned out that among non-Hermitian
Hamiltonians it is possible to allocate a wide number  having real
energy spectra and providing the development of systems over time
with preserving unitarity.

 Now it is well-known fact, that the reality of the
spectrum in models with a non-Hermitian Hamiltonian is a
consequence of $\cal PT$-invariance of the theory, i.e. a
combination of spatial and temporary parity of the total
Hamiltonian: $[H,{\cal PT}]\psi =0$. When the $\cal PT$ symmetry
is unbroken, the spectrum of the quantum theory will be  real.
This surprising results explain the growing interest in this
problem which was initiated by Bender and Boettcher's observation
\cite{ben}. For the past a few years has been studied a lot of new
non-Hermitian $\cal PT$-invariant systems (see, for example
\cite{ft12} - \cite{cmb}).

The algebraic non-Hermitian ${\cal PT}$-symmetric
$\gamma_5$-extension of the Dirac equation was first studied in
\cite{ft12} and further was developed  in the works
\cite{Rod1}-\cite{RodKr1}. However in the geometrical approach to
the construction of  Quantum Field Theory (QFT) with fundamental
mass which was developed by V.G.Kadyshevsky, equation for motion
fermion with $\gamma_5$-mass extension \cite{Kad1} was obtained
yet in seventies years of the last century (see also
\cite{KMRS},\cite{Max}). The purpose of the present paper is the
continuation of the studying examples of pseudo-Hermitian
relativistic Hamiltonians, investigations of which was started by
us earlier (see \cite{Rod1}-\cite{RodKr1}). In the papers
\cite{ROD1},\cite{ROD2}    the  energy spectra of the fermions was
obtained by us as exact solutions of the modified Dirac equation
in which taken into account the interaction of anomalous magnetic
moment (AMM) of fermions with  uniform magnetic field.


On the other hand in 1965  M.A.Markov \cite{Mar} has proposed
hypothesis   according to which the mass spectrum of particles
should be limited by "the Planck mass" $m_{Planck} = 10^{19} GeV
$.  The particles with the limiting mass \be\label{Markov}m \leq
m_{Planck}\ee were named by the author "Maximons". However,
condition (\ref{Markov}) initially was purely phenomenological and
until recently it has seemed that this restriction  can be applied
without connection with SM.  And really SM is irreproachable
scheme for value of mass from zero till infinity. But in the
current situation, however, more and more data are accumulated
that bear witness in favor of the necessity of revising some
physical principles. In particular, this is confirmed by abundant
evidence that "dark matter", apparently  exists and absorbs a
substantial part of the energy density in the Universe.

In the late 1970s, a new radical approach was offered by V. G.
Kadyshevsky \cite{Kad1} (see also \cite{KMRS},\cite{Max}), in
which the Markov's idea of the existence of a maximal mass used as
new fundamental principle construction of QFT. This principle
refutes the affirmation that mass of the elementary particle can
have a value in the interval $0 \leq m < \infty$. In the
geometrical theory the condition finiteness of the mass spectrum
is postulated in the form \be\label{M} m \leq {\cal M},\ee where
the maximal mass parameter ${\cal M}$, was named by the the
\emph{fundamental mass}. This physical parameter is a \emph{ new
physical constant} along at the speed of light and Planck's
constant. The value of ${\cal M}$ is considered as a curvature
radius of a five dimensional hyperboloid whose surface is a
realization of the curved momentum 4-space -- the anti de Sitter
space. Objects with a mass larger than ${\cal M}$ cannot be
regarded as elementary particles because no local fields that
correspond to them. For a free particle, condition (\ref{M}) holds
automatically on surface of a five dimensional hyperboloid. In the
approximation ${\cal M} \gg m$ the anti de Sitter geometry goes
over into the Minkowski geometry in the four dimensional
pseudo-Euclidean space ("flat limit")\cite{Kad1}.

Here we are producing our investigation of non-Hermitian  systems
with $\gamma_5$-mass  contribution taking into account AMM of
fermions in external magnetic field. In Section 2 we consider
restriction of mass in pseudo-Hermitian algebraic theory. We also
are studying the spectral and polarization properties of such
systems (Section 3). The novelty of our approach   is associated
with predictions of new phenomena caused by a number of additional
terms arising in the non-Hermitian Hamiltonians (Section 4).
Intriguing predictions  in our papers \cite{ROD1},\cite{ROD2} are
connected with non-Hermitian mass extension and associated with
the appearance in this the algebraic approach of any new
particles. It is important that previously such particles ("exotic
particles") was observed only in the framework of the geometric
approach to the construction of QFT.


\section{ Restriction of mass in pseudo-Hermitian algebraic theory}

The inequality $m_1 \geq m_2$ in this theory follows from the
condition $m^2={m_1}^2-{m_2}^2$, which is the basic requirement
that defines unbroken symmetry of the Hamiltonian \cite{ft12}.
However, this inequality between $m,\,\,m_1$ and $m_2$ is not
single condition, which links the parameters of
$\gamma_5$-extension of mass. In particular we can write the new
condition for the physical mass $m$, which may be more
substantial. Indeed, using the simple mathematical theorem, we can
obtain inequality in the form \cite{RodKr} \be\label{Alm} m\leq
{{m_1}^2}/{2m_2}= M \ee where the new parameter $M$ restricts
change of mass $m$ (the details of this approach one can see at
Fig.1).

\begin{figure}[h]
\vspace{-0.2cm} \centering
\includegraphics[angle=0, scale=0.5]{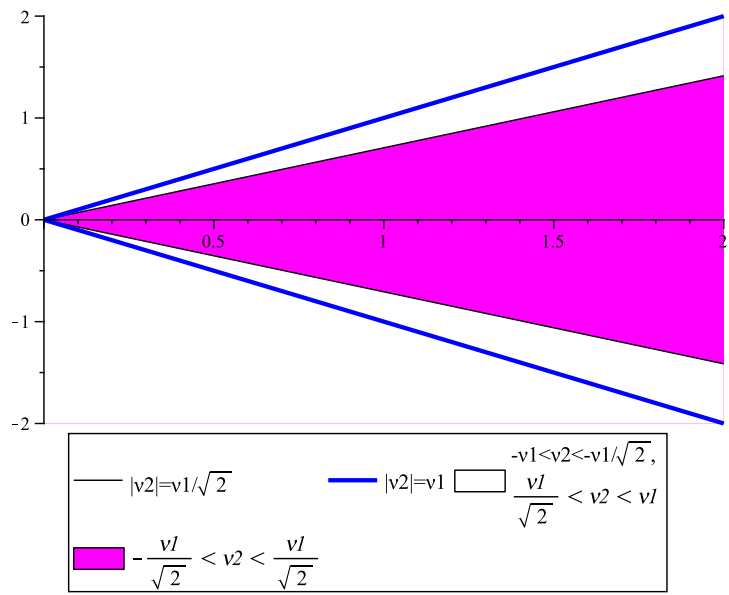}
\caption{The parametric domain of unbroken ${\cal PT}$-symmetry
${m_1}^2\geq {m_2}^2$ for non-Hermitian Hamiltonian comprises
three characteristic sub-domains: the shaded domain II corresponds
to standard particles and the neighboring domains I and III
correspond to the description of exotic fermions.    }
\vspace{-0.1cm}\label{Fig.0-0}
\end{figure}

Introducing the normalized values $\nu=m/M,\, \nu_1 = m_1/M$,
$\nu_2 = m_2/M$ and solving the system of equations $m^2={m_1}^2
-{m_2}^2$ and (\ref{Alm}) we have expressions with two signs for
values parameters $\nu_1$ and $\nu_2$:
\be\label{nu12}{\nu_1}^{\mp}=\sqrt{2(1\mp\sqrt{1-\nu^2})};\,\,\,
\,\,\,\,{\nu_2}^{\mp}=(1\mp\sqrt{1-\nu^2}).\ee
We recall that we are investigating the issue of the existence of
constraints on mass parameters in the given theory. It is shown by
us that there is a constraint on the parameter $m$ in the
algebraic theory.    In this case, there are reasons to believe
that a direct relationship exists between  $M$ obtained by
algebraic way and ${\cal M}$ which is consequence the
\emph{geometric approach to modified QFT with the fundamental
mass} \cite{Rod1},\cite{RodKr}.

Let us now consider obtaining the modified Dirac equations for
free massive particles using the ${\gamma_5}$-factorization of the
ordinary Klein-Gordon operator. It is interesting that in this
case we can use simple way similar to known Dirac's procedure. As
he wrote: "...get something like a square root from the equation
Klein-Gordon" \cite{Dirac1}. And really if we shall not be
restricted to only Hermitian operators then we can represent the
Klein-Gordon operator in the form of a product of two commuting
matrix operators with $\gamma_5$-mass extension (where $\hbar=c
=1$):

\be\label{D2} \Big({\partial_\mu}^2 +m^2\Big)=
\Big(i\partial_\mu\gamma^{\mu}-m_1-\gamma_5 m_2 \Big)
\Big(-i\partial_\mu\gamma^{\mu}-m_1+\gamma_5 m_2 \Big), \ee where
 the physical mass of
particles $m$ is expressed through the parameters $m_1$ and $m_2$
 \be \label{012} m^2={m_1}^2- {m_2}^2. \ee

For  the function would obey to the equations of Klein-Gordon
\be\label{KG} \Big({\partial_\mu}^2
+m^2\Big)\widetilde{\psi}(x,t)=0 \ee one can demand that it also
satisfies to one of equations of the first order \be\label{ModDir}
\Big(i\partial_\mu\gamma^{\mu}-m_1-\gamma_5 m_2
\Big)\widetilde{\psi}(x,t)
=0;\,\,\,\Big(-i\partial_\mu\gamma^{\mu}-m_1+\gamma_5 m_2 \Big)
\widetilde{\psi}(x,t)=0 \ee

Equations (\ref{ModDir}) of course, are less common than
(\ref{KG}), and although every solution of one of the equations
(\ref{ModDir}) satisfies to (\ref{KG}), reverse approval has not
designated.

It is also obvious that the Hamiltonians, associated with the
equations (\ref{ModDir}), are non-Hermitian (pseudo-Hermitian),
because in them the $\gamma_5$-dependent mass components appear
($H\neq H^{+}$):

  \be\label{H} H =\overrightarrow{\alpha} \textbf{p}+ \beta(m_1
+\gamma_5 m_2)=\overrightarrow{\alpha} \textbf{p}+ \beta m
e^{\gamma_5 \alpha} \ee and \be\label{H+} H^+
=\overrightarrow{\alpha }\textbf{p}+ \beta(m_1 -\gamma_5
m_2)=\overrightarrow{\alpha} \textbf{p}+ \beta m e^{-\gamma_5
\alpha}.\ee Here  matrices $\alpha_i=\gamma_0\cdot\gamma_i$,
$\beta=\gamma_0$, $\gamma_5=-i\gamma_0\gamma_1\gamma_2\gamma_3$
and introduced identical replacement of parameters
\be\label{alpha}\sinh(\alpha)=m_2/m;
\,\,\,\,\cosh(\alpha)=m_1/m,\ee where parameter $\alpha$ varies
from zero to infinity.

It is easy to see from (\ref{012}) that the  mass $m$, appearing
in the equation (\ref{KG}) is real, when the inequality \be
\label{e210} {m_1}^2\geq {m_2}^2,\ee is accomplished\cite{ft12}.
However for variable $\alpha$ which is identical for definitions
$m_1,\,\,\,m_2$ this  condition is automatically accomplished in
all region  of $\alpha$ changes.


However this area contains descriptions not only pseudo-Hermitian
fermions, which in a result of transition to Hermitian limit ($m_2
\rightarrow 0, \,\, m_1\rightarrow m$)  coincide with the ordinary
particles (see Fig.\ref{Fig.0-0} and \cite{ft12}). But  there are
the second region where fermions do not subordinate to the
ordinary Dirac equations and for them the Hermitian transition is
absent. It is easy to see that (\ref{012}) can be used for new
restrictions of mass parameters. Really, if we take into account
inequality between arithmetic and geometrical averages of two
positive numbers we have

\be\label{m2M} m^2 + {m_2}^2 \geq 2\sqrt{m^2\,{m_2}^2 }.\ee \vskip
0.5cm

On the foundation of this inequality  it can formulate five
important Remarks.

\vskip 0.5cm

 \textbf{ Remark 1.} \emph{The pseudo-Hermitian  approach with $\gamma_5$-mass extension
 contains restriction of mass parameter beside (\ref{012}).}  Indeed from (\ref{m2M}) follows,
 that the sign of equality takes place   when $m = m_2$. If we use
 parameter $\alpha $ then from (\ref{m2M})  \be\label{alpha}
 1+\sinh^2\alpha = 2 \sinh\alpha. \ee From (\ref{alpha})
 we can also see that in this point
 $\sinh\alpha =1$ and solving this equation we can  find
 the  value:  $$\alpha_0 = 0.881 $$.

 \textbf{ Remark 2.} \emph{Maximal value of mass particle $m=M$ is  achieved
 in the point $m_2=M\,\,\,m_1=\sqrt{2}M$. }The proof
 of this fact can be confirmed by the way defining the mass of the Maximon.
 Indeed,  under  $\alpha_0$ we have the equality $m\sinh\alpha_0=M= m_2$.
and also for $m_1$ we have  $m_1= m \cosh\alpha_0=\sqrt{2}M$.

 \textbf{  Remark  3.}   \emph{The particle with the maximal mass (Maximon) is
 the pseudo-Hermitian fermion.} Using Remark 2. for Maximon  we
    can obtain  expression  \be\label{Max}  M_{Maximon}=\sqrt{2}M+\gamma_5 M. \ee
    This phenomenon  may be given a very simple physical
    interpretation.
  This means that particles with the maximal mass (Maximons) $m=M$ are non-Hermitian
  (pseudo-Hermitian) fermions.

 \begin{figure}[h]
\vspace{-0.2cm} \centering
\includegraphics[angle=0, scale=0.5]{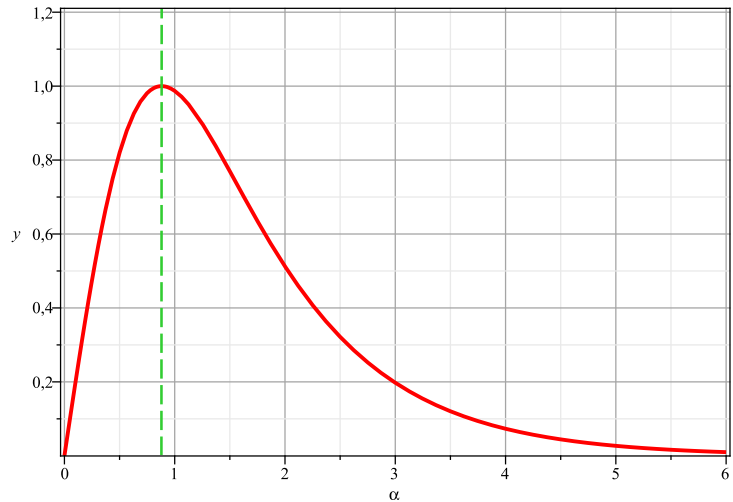}
\caption{Dependence of $y(\alpha)=\tilde{m} = m/M $ on the
parameter $\alpha$} \vspace{-0.1cm}\label{Fig.0-5}
\end{figure}

At the Fig.\ref{Fig.0-5} we can see explicit behavior of the
reduced mass distribution of $y(\alpha)=m/M$, depending on
parameter $\alpha.$ From this picture follows that the curve,
corresponding mass of the considered particles, has a maximum.
This the maximal value of $y(\alpha)= m/M = 1 $ corresponds to
$\alpha_0 =0.881$ that as already noted correspond to Maximon.
Till to this value we are dealing with fermions, which have
Hermitian limit when $M\rightarrow \infty$ (or $m_2\rightarrow
0$). But after the value $\alpha_0 =0.881$ is achieved we once
more deal with decreasing mass of particles. However in this
region already no the possibility with the help of the limiting
transition to obtain Hermitian mass. Thus in this region particles
exist, which in principal differ from the particles of the SM
(exotic particles).

\textbf{ Remark 4. } This particles (exotic particles) exist
thanks to the presence of a maximal value of mass parameter,
because the Hermitian limit for them  is absent.  If one can
detect their it means that limiting mass fermions exist and our
world is pseudo-Hermitian.

\begin{figure}[h]
\vspace{-0.2cm} \centering
\includegraphics[angle=0, scale=0.5]{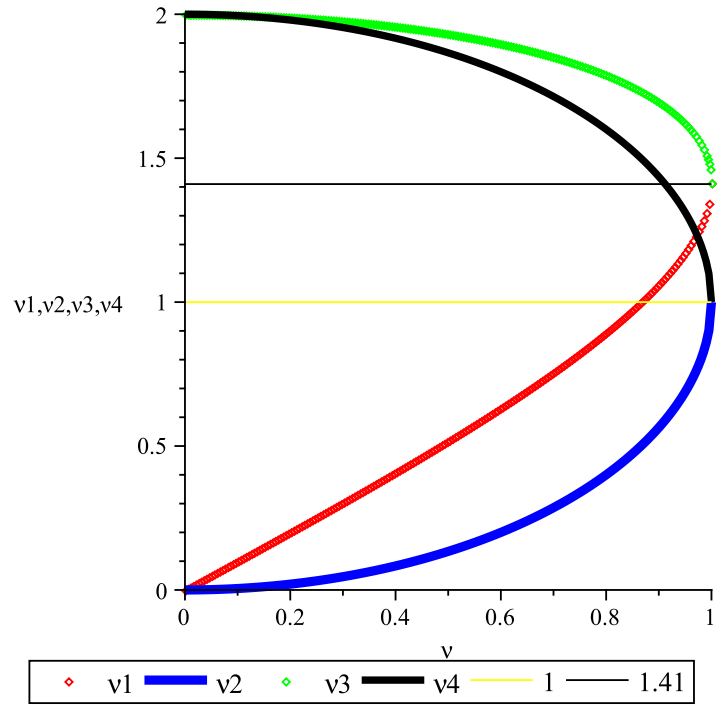}
\caption{Values of the parameters
$\nu_1={m_1}^{-}/M,\nu_2={m_2}^{-}/M,\nu_3={m_1}^{+}/M,
\nu_4={m_2}^{+}/M $  as functions of $\nu=m/M $ }
\vspace{-0.1cm}\label{Fig.1-2}
\end{figure}

\textbf{ Remark 5. }  And vice versa if the restriction of mass
spectrum of elementary particles does not exist  then exotic
particles can not arise in Nature. And it is very important
because   restriction of mass in SM is absent  then experimental
verification can be start from the most biggest values of maximal
mass. In particular,  it may be the Planck mass $M = 10^{19} GeV$.

 It is very interesting that the early such particles had discovered
 in geometrical approach to the construction of QFT with fundamental
 mass \cite{Kad1}-\cite{Max}.  We believe that exact solutions of
 modified Dirac-Pauli equations
which were obtained by us for the pseudo-Hermitian neutrinos can
let valuable information to detect the presence of "exotic
particles". This is indicated also increase the effects
proportionally $\sim M/m_{\nu} $ associated with unusual
properties of "exotic neutrinos", interacting with the magnetic
fields.

At Fig.\ref{Fig.1-2} we can see values of different branches of
${\nu_1}^\pm$ and ${\nu_2}^\pm$  as a function of normalized
physical parameter $\nu$ (see also (\ref{nu12})). The existence
the domain of the ${\cal PT}$-symmetry is $0\leq \nu \leq 1$. For
these values of the parameters $\nu_1$ and $\nu_2$, the modified
Dirac equation with the maximum mass describes the propagation of
particles with real masses. But the lower branches ${\nu_1}^- $,
${\nu_2}^- $ correspond to ordinary particles and upper lines
${\nu_1}^+$, ${\nu_2}^+$ define  the exotic  partners.





\section{ Modified model for the study of non-Hermi-tian mass parameters in intensive magnetic fields}

In this section, we shell  want touch upon question of describing
the motion of Dirac particles, if their own magnetic moment is
different from the Bohr magneton. As it was shown by Schwinger
\cite{Sc}  the equation of Dirac particles in the external
electromagnetic field $A^{ext}$ taking into account the radiative
corrections may be represented in the form: \be\label{A}
\left({\cal P}\gamma - m\right)\Psi(x)-\int{\cal
M}(x,y|A^{ext})\Psi(y)dy=0, \ee where ${\cal M}(x,y|A^{ext})$ is
the mass operator of the fermion in the external field and ${\cal
P}_\mu =p_\mu - {A^{ext}}_\mu$ . From equation (\ref{A}) by means
of expansion of the mass operator in a series of according to $
eA^{ext}$ with precision not over then linear field terms one can
obtain the modified equation. This equation preserves the
relativistic covariance and consistent with the phenomenological
equation of Pauli obtained in his early papers (see for example
\cite{TKR}).

Now let us consider the model of massive fermions with
$\gamma_5$-extension of mass $m\rightarrow m_1+\gamma_5 m_2$
taking into account the interaction of their charges and AMM with
the electromagnetic field $F_{\mu\nu}$:

\be\label{Delta} \left( \gamma^\mu {\cal P}_\mu -
 m_1 -\gamma_5 m_2 -\frac{\Delta\mu}{2}\sigma^{\mu \nu}F_{\mu\nu}\right)\widetilde{\Psi}(x)=0,\ee
where $\Delta\mu = (\mu-\mu_0)= \mu_0(g-2)/2$. Here $\mu$ -
magnetic moment of a fermion, $g$ - fermion gyromagnetic factor,
$\mu_0=|e|/2m$ - the Bohr magneton,
$\sigma^{\mu\nu}=i/2(\gamma^\mu \gamma^\nu-\gamma^\nu
\gamma^\mu)$. Thus phenomenological constant $\Delta\mu $, which
was introduced by Pauli,  is part of the equation and gets the
interpretation with the point of view QFT.

The Hamiltonian form of (\ref{Delta}) in the homogenies magnetic
field is the following \be i\frac{\partial}{\partial t}
\widetilde{\Psi}(r,t)=H_{\Delta \mu}\widetilde{\Psi}(r,t),\ee
where \be\label{Delta1} H_{\Delta\mu} = \vec{\alpha}\vec{{\cal P}}
+ \beta(m_1 + \gamma_5 m_2) +
\Delta\mu\beta(\vec{\sigma}\textbf{H}).\ee For example, given the
quantum electrodynamic contribution in AMM of an electron with
accuracy up to $e^2$ order we have
$\Delta\mu=\frac{\alpha}{2\pi}\mu_0 $, where $\alpha = e^2 =1/137$
- the fine-structure constant and we still believe that the
potential of an external field satisfies to the free Maxwell
equations.

It should be noted that now the operator projection of the fermion
spin at the direction of  its movement  - $\overrightarrow{
\sigma} \overrightarrow{{\cal P }} $ is not commute with the
Hamiltonian (\ref{Delta1}) and hence it is not the integral of
motion. The operator  $\mu_3$  commuting with this Hamiltonian  is
the operator of polarization which can be represented in the form
of the third component of the polarization tensor is oriented
along  the direction of the magnetic field  \cite{TKR}
\be\label{muH}
          \mu_3=m_1\sigma_3 + \rho_2[\vec{\sigma}\vec{{\cal
          P}}]_3
\ee where matrices $$ \sigma_3= \left(%
\begin{array}{cc}
  I & 0 \\
  0 & -I \\
\end{array}%
\right); \,\,\,\,\,              \rho_2 = \left(%
                                      \begin{array}{cc}
                                        0 & -iI \\
                                     iI & 0 \\
                                         \end{array}%
                                       \right).
$$
Subjecting the wave function $\widetilde{ \psi }$ to requirement
to be eigenfunction of the operator polarization $\mu_3$ and
Hamilton operator (\ref{Delta1}) we can obtain: \be\label{Pi}
\mu_3\widetilde{\psi} = \zeta k\widetilde{\psi}, \,\,\, \mu_3=\left(%
\begin{array}{cccc}
  m_1 & 0 & 0 & {\cal P}_1-i{\cal P}_2 \\
  0 & -m_1 & -{\cal P}_1-i{\cal P}_2 & 0 \\
  0 & -{\cal P}_1+i{\cal P}_2 & m_1 & 0 \\
  {\cal P}_1+i{\cal P}_2 & 0 & 0 & -m_1 \\
\end{array}%
\right), \ee where $\zeta=\pm 1$ are characterized the projection
of fermion spin at the direction of the magnetic field, and
 $$H_{\Delta\mu}\widetilde{\psi}=E\widetilde{\psi},$$
  \be\label{Hmu} H_{\Delta\mu}=\left(%
\begin{array}{cccc}
  m_1+H\Delta\mu & 0 & {\cal P}_3 -m_2& {\cal P}_1-i{\cal P}_2 \\
  0 & m_1-H\Delta\mu & {\cal P}_1+i{\cal P}_2  & -m_2-{\cal P}_3\\
  m_2+{\cal P}_3 & {\cal P}_1-i{\cal P}_2 & -m_1-H\Delta\mu & 0 \\
  {\cal P}_1+i{\cal P}_2 & m_2-{\cal P}_3 & 0 & H\Delta\mu-m_1 \\
\end{array}%
\right). \ee

\section{ Exact solutions of Dirac-Pauli equations in the intensive
uniform  magnetic field}

Performing calculations  in many ways reminiscent of similar
calculations carried out in the ordinary model in the magnetic
field \cite{TKR} - \cite{3}, in a result, for modified Dirac-Pauli
equation one can find \emph{the exact solution for energy
spectrum} \cite{ROD1},\cite{ROD2}:
 \be\label{E61} E(\zeta,p_3,2\gamma
n,H)=\sqrt{{p_3}^2-{m_2}^2+\left[\sqrt{{m_1}^2+2\gamma
n}+\zeta\Delta\mu H \right]^2} \ee and for eigenvalues of the
operator polarization $\mu_3$ we can write in the form \be
k=\sqrt{{m_1}^2 +2\gamma n}. \ee

From (\ref{E61}) it follows that in the field  where  ${\cal PT}$
symmetry is unbroken $m \leq M$, all energy levels are real for
the case of spin orientation along the magnetic field direction
$\zeta =+1$.

\begin{figure}[h]
\vspace{-0.2cm} \centering
\includegraphics[angle=0, scale=0.5]{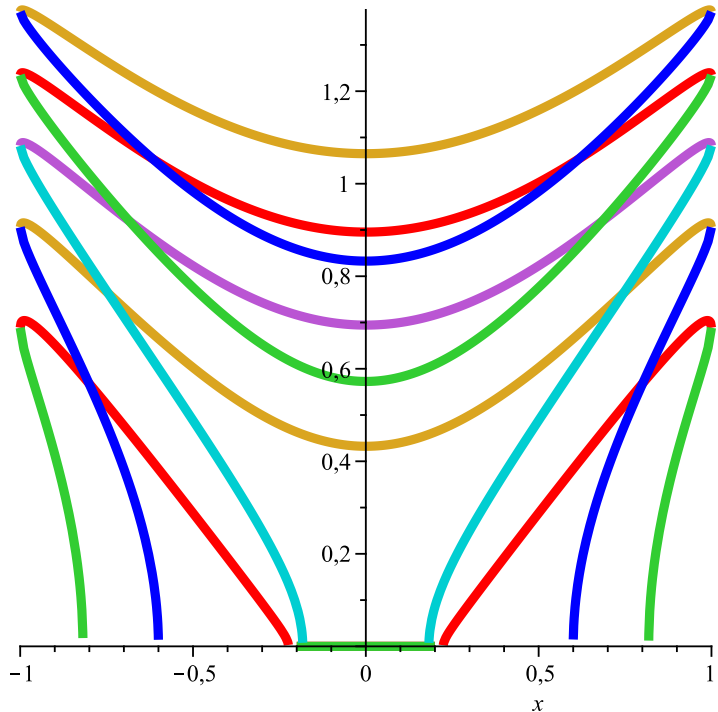}
\caption{Dependence of $E(-1,0, 0.4n, 0.1 )$ on the parameter
$x=m/M$ for the cases $n=0,1,2,3,4\,\, and\,\,\Delta\mu H = 0.1.$}
\vspace{-0.1cm}\label{Fig5-1}
\end{figure}

However, in the opposite case $\zeta = -1$ we have the  imaginary
part from the ground state of fermion $n=0$ and other low energy
levels, see on Fig.\ref{Fig5-1}. For the cases of increasing
parameter $\Delta\mu H = 0.2$ we can watch overlapping of
different levels.

It is easy to see that in the case $ \Delta\mu =0$ from
(\ref{E61}) one can obtain the ordinary expression for energy of
charged particle in the magnetic field (\emph{Landau levels}).
Besides it should be emphasized that from the expression
(\ref{E61}), in the Hermitian limit putting $m_2=0$ and $m_1=m$
one can obtain: \be\label{DPA} E(\zeta,p_3,2\gamma
n,H)=\sqrt{{p_3}^2+\left[\sqrt{{m}^2+2\gamma n}+\zeta\Delta\mu H
\right]^2}.
 \ee

Note that in the paper \cite{TBZ} was previously obtained result
analogical to  (\ref{DPA}) by means of using of the Hermitian
Dirac-Pauli approach. Direct comparison of formula (\ref{DPA})
with the  result \cite{TBZ} shows their coincidence in the
Hermitian limit $M\rightarrow \infty$. It is easy to see that the
expression (\ref{E61}) contains dependence on parameters $m_1$ and
$m_2$ separately, which are not combined into \emph{a mass of
particles}, that essentially differs from the examples which were
considered early \cite{ft12}-\cite{RodKr}.

 Thus, here the calculation of
interaction AMM of fermions with the magnetic field allow to put
the question about the possibility of experimental studies of the
non-Hermitian effects of $\gamma_5$-extensions of a fermion mass.
Thus, taking into account the expressions (\ref{E61}) we obtain
that the energetic spectrum  is expressed through the fermion mass
$m$ and the value of the maximal mass $M$. Thus, taking into
account that the interaction AMM with magnetic field removes the
degeneracy on spin variable, we can obtain the energy of the
ground state ($\zeta=-1$) in the form which dependence  is
represented at Fig.\ref{Fig5-1}.

Thus, it is shown that the main progress, is obtained by us in the
algebraic way of the construction of the fermion model with
$\gamma_5$-mass term is consists of describing of the new
energetic scale, which is defined by the parameter
$M={m_1}^2/2m_2$. This value on the scale of the masses is a point
of transition from the ordinary particles $m_2 < M$ to exotic $m_2
> M $. Furthermore, description of the exotic fermions in the
algebraic approach are turned out essentially the same as in the
model with a maximal mass, which was investigated by
V.G.Kadyshevsky with colleagues on the basis of geometrical
approach \cite{Kad1}-\cite{Max}.

It should be noted that the formula (\ref{E61})  is a valid not
only for charged fermions, but and for the neutral particles
possessing AMM. In this case one must simply replace the value of
quantized transverse momentum of a charged particle in a magnetic
field on the ordinary value $2\gamma n\rightarrow
{p_1}^2+{p_2}^2$.
 Thus, for the case of ultra cold polarized ordinary electronic
 neutrino with precision not over then linear
field terms   we can write

\be\label{E34} E(-1,0,0,H,M \rightarrow \infty)= m_{\nu_e}
\sqrt{1-\frac{\mu_{\nu_e}}{\mu_0}\frac{
 H}{ H_c}}.
 \ee
However, in the case of exotic electronic
 neutrino $\tilde{m}_{\nu_e} $ we have
 \be\label{E3}
E(-1,0,0,H,\tilde{m}_{\nu_e}/M)= \tilde{m}_{\nu_e}
\sqrt{1-\frac{\mu_{\nu_e}}{\mu_0}\frac{2 M
 H}{{\tilde{m}_{\nu_e}} H_c}}.
 \ee

It is well known \cite{n3},\cite{n33} that in the minimally
extended SM the one-loop radiative correction generates neutrino
magnetic moment which is proportional to the neutrino mass
\be\label{mu1}
  \mu_{\nu_e}=\frac{3}{8\sqrt{2}\pi^2}|e| G_F
  m_{\nu_e}=\left(3\cdot10^{-19}\right)\mu_0\left(\frac{m_{\nu_e}}{1
  eV}\right),
\ee where $ G_F$-Fermi coupling constant and $\mu_0$ is Bohr
magneton.
 However note
that the best laboratory upper limit on a neutrino magnetic
moment, $\mu\leq 2.9 10^{-11}\mu_0$, has been obtained by the
GEMMA collaboration \cite{mu1}, and the best astrophysical limit
is $\mu\leq 3\cdot10^{-12} \mu_0$.

 Existence masses of
neutrino and mixing implies that neutrinos have magnetic moments.
In last time one can often meet with  an overviews   of
electromagnetic properties neutrino, (see, for example,
\cite{Ternov1}). But as it was noted up in this paper " now there
is no positive experimental indication in favor existence
electromagnetic properties of neutrinos".  With it really is hard
not to agree because  the interactions of ordinary neutrinos with
the electromagnetic fields are extremely weak. However if one to
suggest using the "exotic neutrinos" the interaction with magnetic
field  may be really significantly increased thanks to the
coefficient which be equal to the ratio of maximal mass and mass
of   neutrinos $k=M/m_{\nu}$\cite{ROD1},\cite{ROD2}. Such
experiments in our opinion may  be very fruitful for creation  the
new physics beyond the SM. Perhaps that  this effects  indeed can
be observed in terrestrial experiments.

 \end{document}